\documentclass[twocolumn]{aastex62}
\pdfoutput=1

\graphicspath{{./}{figures/}}
\usepackage{natbib}
\usepackage{hyperref}

\received{November 27, 2019}
\revised{May 14, 2020}
\accepted{May 14, 2020}

\submitjournal{JAAVSO}

\shorttitle{AG~Dra 2018 Disk Instability Outburst}
\shortauthors{Richie}

\begin{document}

\title{Disk Instabilities Caused the 2018 Outburst of AG~Draconis}

\author{Helena M. Richie}
\author{W. M. Wood-Vasey}
\author{Lou Coban}
\affil{Physics and Astronomy Department, University of Pittsburgh, 
3941 O'Hara St, Pittsburgh, PA 15260}

\begin{abstract}
Symbiotic binary AG~Draconis (AG~Dra) has an well-established outburst behavior
based on an extensive observational history. Usually, the system undergoes a 9--15~yr period of quiescence with a constant average energy emitted, during which the system's orbital period of $\sim$550~d can be seen at shorter wavelengths (particularly in the U-band) as well as a shorter period of $\sim$355~d thought to be due to pulsations of the cool component. After a quiescent period, the marker of an active period is usually a major (cool) outburst of up to $\textrm{V}=8.4$~mag, followed by a series of minor (hot) outbursts repeating at a period of approximately 1~yr. However, in 2016 April after a 9-year period of quiescence AG~Dra exhibited unusual behavior: it began an active phase with a minor outburst followed by two more minor outbursts repeating at an interval of $\sim$1~yr. We present R-band observations of AG~Dra's 2018 April minor outburst and an analysis of the outburst mechanism and reports on the system's activity levels following the time of its next expected outburst. By considering the brightening and cooling times, the scale of the outburst, and its temperature evolution we have determined that this outburst was of disk instability nature.
\end{abstract}

\keywords{symbiotics, AG~Draconis, AAVSO}

\section{Background} \label{sec:bkgd}
\subsection{Symbiotics} \label{sec:symbiotics}
Symbiotic binary systems, also known as symbiotics, are a type of cataclysmic variable star (CV) that consist of an interacting cool giant star and a hot compact object, most commonly a white dwarf (WD). Interaction between the cool and hot component results from an outflow of matter from the cool component that accretes onto the hot component. Matter outflow can be due to stellar wind off of the cool component or the cool component overfilling its Roche-lobe. In many cases, the rate of mass loss off of the cool component can be sufficient to fuel hydrogen burning in a thermonuclear shell around the hot component \citep[][and references therein]{Sokoloski_2006}. As a result of mass outflow, symbiotics often exist inside of a gas cloud that can be fully or partially ionized by the hot component \citep{1986syst.book.....K}.

Symbiotics undergo periods of quiescence and activity, driven by the state of equilibrium between mass outflow, accretion, and ionization of the gas cloud. In quiescence, symbiotics emit energy at a constant average rate. During an active phase where this equilibrium is disturbed, symbiotics can be observed to undergo outbursts that feature an optical brightening of the system by 2--3 magnitudes with amplitude decreasing at longer wavelengths \citep{2003ASPC..303..249S}. This classical symbiotic outburst (or classical novae) is the most common type of outburst and commonly recurs on timescales of a decade \citep{1986syst.book.....K}. The driving mechanism behind a classical symbiotic outburst is the shedding of material off of the cool component onto the hot component as it overfills its Roche lobe, triggering thermonuclear runaway in a shell on the surface of the hot component. Another common type of outburst observed in CV systems containing red dwarfs (as opposed to giants) is the dwarf nova, which is driven by instability in accretion disks surrounding the WD that cause an increase in mass flow through the disk, resulting in temporary heating and brightening. These types of outbursts necessarily have smaller peak magnitudes and timescales than those observed in classical symbiotic outbursts.

Though these mechanisms are reasonable explanations of some outbursts observed in symbiotics, there are many outbursts that have been observed that, due to their the scales and recurrence times, cannot be explained by these mechanisms alone \citep{1986syst.book.....K,1995AJ....109.1289M,1979ApJ...230..832S}. To this end, \citet{Sokoloski_2006} proposed a \emph{combination nova} model to describe outbursts in symbiotics that exhibit qualities of both classical symbiotic outbursts and dwarf novae. This model suggests that outbursts in symbiotics are due to enhanced thermonuclear burning with disk instability as a trigger event. This model can account for the peak luminosities and short recurrence times of outbursts seen in many symbiotics, particularly for Z~Andromedae as described in \citet{Sokoloski_2006}.

On a larger scale, studying symbiotics is important in the context of their being a possible progenitor of Type Ia supernovae. As such, we would like to understand the true nature of these outbursts in order to predict their activity patterns. For most symbiotics, we have not been able to observe state transitions from quiescence to activity due to their irregular outburst behavior. An interesting exception is the system known as AG~Draconis (AG~Dra). This symbiotic has cyclical activity patterns, making it possible to predict and observe its state transitions.

\subsection{AG~Draconis} \label{sec:agdra}
AG~Dra is one of the best-studied symbiotics, with observations spanning the last century. Like most symbiotics, AG~Dra has been observed to alternate between phases of quiescence and activity, undergoing a series of outbursts during its active phases. According to \citet{GR1999}, such outbursts can be of both hot and cool type. Cool outbursts are caused by the expansion of the hot component's pseudo-atmosphere and a subsequent drop in WD temperature, which can be seen as an anticorrelation between optical/UV and X–ray emission. Hot outbursts occur when the WD's radius remains fixed and its temperature increases or remains the same. Outbursts of this nature show consistencies with disk instability-driven dwarf novae outbursts, as well as with the \citet{Sokoloski_2006} combination nova model where the thermonuclear burning pseudo-atmosphere of the WD expands after exceeding a threshold accretion rate triggered by disk instabilities. Evidence of the existence of an accretion disk surrounding the WD has recently been provided in a study done by \citet{2019MNRAS.487.2166L}. Over the course of its observation, certain periods have been discovered that characterize the system's orbital motion and outburst behavior. With its semi-regular state transitions, AG~Dra is a useful subject to study in order to characterize the mechanisms of symbiotic outbursts that are generally unclear.

The system consists of a K3 III red giant \citep{1987AJ.....93..938K} and white dwarf that are 1.5~$M_\odot$ and 0.4--0.6~$M_\odot$ \citep{1995AJ....109.1289M}, respectively. The hot component has a luminosity of $\sim10^3~L_\odot$ and a temperature of $\sim80-150\times10^3~K$ \citep{1995AJ....109.1289M}. The components have been observed to be at an orbital separation of 400~$R_\odot$ \citep{1986AJ.....91.1400G} and are enveloped in a partially ionized circumbinary nebula \citep{1995AJ....109.1289M}. \citet{2019MNRAS.487.2166L} showed that the upper limit in accretion disk size is $0.3~\textrm{au}$ or $\sim65~R_\odot$. Radio observations of emission from the circumbinary nebula give a rate of mass loss of $10^{-7}~M_\odot \textrm{yr}^{-1}$ \citep{1995AJ....109.1289M}. There is also evidence of thermonuclear shell burning on the WD's surface at a rate of $3.2\times10^{-8}~M_\odot \textrm{yr}^{-1}$ \citep{GR1999}. From a study of the historical UBV light curve of AG~Dra done by \citet{Hric2014}, the time between active periods has been observed to be anywhere from 12--16~yr. Additionally, two periods for the system have been clearly established; an orbital period of $\sim$550~d and a period of $\sim$355~d thought to be a result of pulsations of the cool component \citep{Galis1999}. AG~Dra's orbital period becomes prominent at shorter wavelengths, showing itself most clearly in the U-band during quiescence. Its pulsation period can be seen during both quiescent and active phases and is most visible in B and V-bands \citep{Galis2016}.

In its observed active phases, AG~Dra exhibits outbursts with consistent peak magnitudes, but irregular multitudes and shapes. They are spaced anywhere from 359--375~d \citep{2015OEJV..169....4G}. Normally, after an extended period of quiescence at $\textrm{V}=9.8$~mag, AG~Dra begins its active phases with a major cool outburst with peak magnitude of about $\textrm{B}=8.8$~mag and $\textrm{V}=8.4$~mag \citep{Galis2017} followed by a series of minor hot outbursts. It has not been confirmed, but the combination nova outburst model seems like a promising explanation of the underlying mechanism for a number of these outbursts. However, in May of 2015 AG~Dra exhibited very unusual behavior as it entered its most recent active phase. The activity began with a minor outburst with peak magnitude of $\textrm{V}=9.6$~mag, followed (at the usual cadence of $\sim$360~d) by two more minor outbursts with peak magnitude of $\textrm{B}=9.1$~mag and $\textrm{V}=9.6$~mag \citep{Galis2017}. This study showed that during these minor outbursts the system exhibited signs of both hot and cool type outbursts by examining the equivalent widths of certain emission lines and observing the disappearance of the Raman scattered O VI lines, respectively. In early April of 2018, AG~Dra began its fourth minor outburst of its 2015--2018 active phase. The Survey of Transiting Extrasolar Planets at the University of Pittsburgh (STEPUP) has monitored this outburst by conducting R-band photometric observations to examine an understudied band-pass of the system's outbursts. With these measurements, we seek to characterize the nature of AG~Dra's most recent outburst.
%%%%%%%%%%%%%%%%%%%%%%%%%%%%%%%%%%%%%%%%%%%%%%%%%%%%%%%%%%%%

\begin{table*}[t]
  \centering
  \caption{ Sample first ten data points of SIA output for STEPUP AG~Dra observations.}
  \begin{tabular}{ccccccccc}
    \hline
    \hline
    Date & Date & ExpTime & Target R & Uncertainty & Filter & Check Label & Check R & Airmass \\
     & [JD] & [s] & [mag] & [mag] & & & [mag] \\
    \hline
    2018-04-30 & 2458239.611990740 & 30 & 8.5960 & 0.0008672 & R & 345 & 11.8312 & 1.3478 \\
    2018-04-30 & 2458239.616747690 & 30 & 8.5818 & 0.0008344 & R & 345 & 11.6420 & 1.3338 \\
    2018-04-30 & 2458239.623541670 & 30 & 8.5725 & 0.0008411 & R & 345 & 11.6270 & 1.3148 \\
    2018-04-30 & 2458239.624884260 & 30 & 8.5966 & 0.0008593 & R & 345 & 11.6529 & 1.3111 \\
    2018-04-30 & 2458239.625578700 & 30 & 8.5749 & 0.0008360 & R & 345 & 11.6265 & 1.3093 \\
    2018-04-30 & 2458239.626250000 & 30 & 8.5678 & 0.0008362 & R & 345 & 11.6211 & 1.3075 \\
    2018-04-30 & 2458239.627615740 & 30 & 8.5854 & 0.0008440 & R & 345 & 11.6348 & 1.3038 \\
    2018-04-30 & 2458239.628287040 & 30 & 8.5789 & 0.0008348 & R & 345 & 11.6345 & 1.3020 \\
    2018-04-30 & 2458239.629641200 & 30 & 8.5690 & 0.0008288 & R & 345 & 11.6299 & 1.2985 \\
    2018-04-30 & 2458239.630324070 & 30 & 8.5812 & 0.0008463 & R & 345 & 11.6208 & 1.2968 \\
    ... & ... & ... & ... & ... & ... & ... & ... & ... \\
    \\
  \hline
  \end{tabular}
 \label{tab:siatable}
 \tablecomments{First ten data points from STEPUP’s observations of AG Dra’s2018 outburst. The full table is available through the AAVSO ftp site at
\href{ftp://ftp.aavso.org/public/datasets/richie481-sia-agdra-output.txt}{ftp://ftp.aavso.org/public/datasets/richie481-sia-agdra-output.txt} (if necessary, copy and paste link into browser).}
\end{table*}

\section{STEPUP} \label{sec:stepup}

STEPUP\footnote{\href{http://pitt.edu/~stepup/}{www.pitt.edu/$\sim$stepup}} has used the Meade Instruments f/8, 16'' RCX400 Keeler Telescope at  the Allegheny Observatory in Pittsburgh, Pennsylvania, USA to conduct photometric observations of a variety of objects since its inception in 2009 \citep{pittir9073}. The main camera is a Santa Barbara Instruments Group (SBIG) STL-6303e and the field of view is 29.2 arcminutes by 19.5 arcminutes. Founded by Melanie Good, STEPUP's original mission was to discover and study new transiting exoplanets and has recently expanded its reach to observing variable stars. STEPUP records their data and processes it with an image analysis program, STEPUP Image Analysis (SIA)\footnote{\href{https://github.com/mwvgroup/STEPUP_image_analysis}{www.github.com/mwvgroup/STEPUP\_image\_analysis}}, written in the Python programing language by lead undergraduate, Helena Richie. SIA is responsible for removing instrument signatures from STEPUP's data, generating WCS information for each file, and performing differential aperture photometry to generate a light curve of the target object. SIA uses the AstroPy Python package \citep{astropy:2018} throughout the routine as well as the WCSTools \citep{1997ASPC..125..249M} software package and Astrometry.net \citep{2010AJ....139.1782L} in the process of plate-solving the images. STEPUP has contributed to several publications \citep{Shporer2010, Fleming_2012} on exoplanet transit timing variations and discovery.
%%%%%%%%%%%%%%%%%%%%%%%%%%%%%%%%%%%%%%%%%%%%%%%%%%%%%%%%%%%%

\begin{figure*}
\centering
\textbf{STEPUP Observations of AG~Draconis 2018 Outburst}\par\medskip
\plotone{{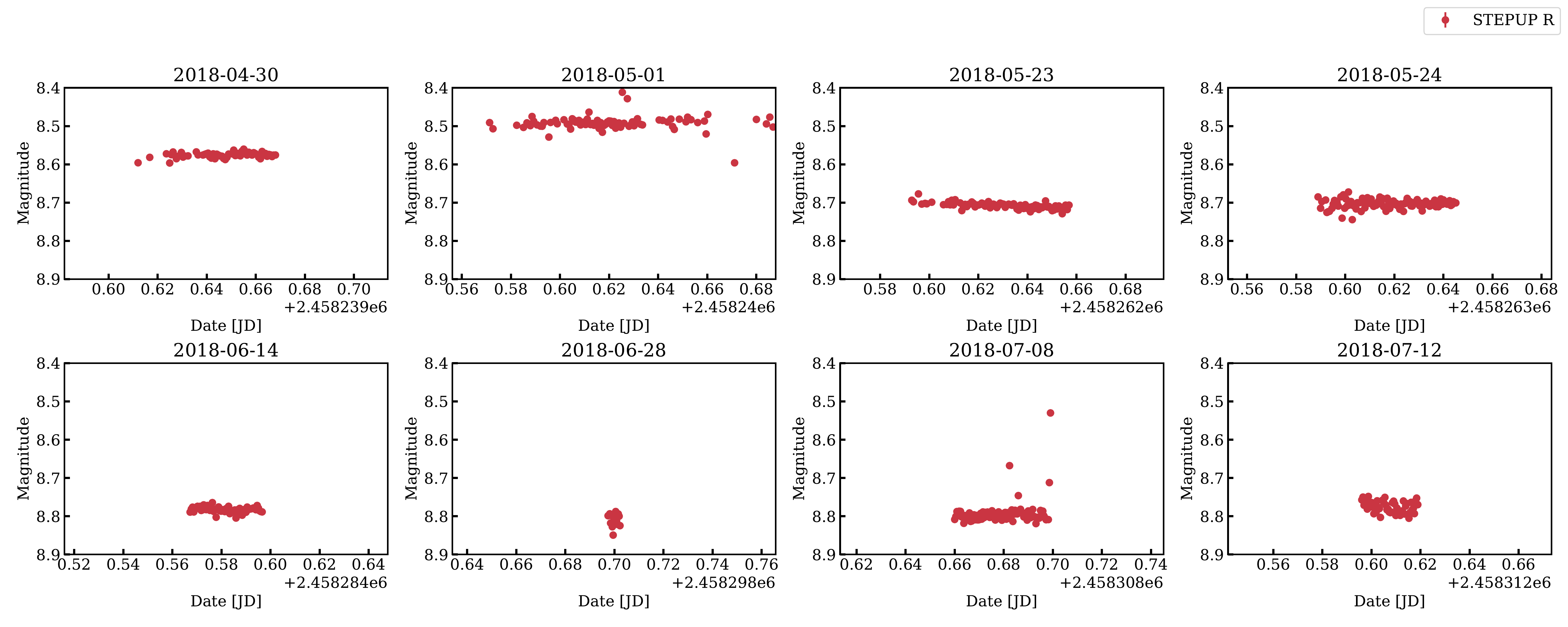}}
\caption{STEPUP observations of AG~Draconis 2018 outburst. Figures are shown on a $0.13183~\textrm{d}=3.1639~\textrm{h}$ timescale. These observations show no obvious variation in brightness and thus put a lower limit on the brightening timescale of AG~Dra during outburst. \label{fig:brightening}}
\end{figure*}

\section{Observations} \label{sec:obs}

To monitor AG~Dra's outburst behavior, STEPUP began conducting observations of the system in late 2018 April and continued through 2018 July using the Cousins R filter.  Observations were made using a variety of exposure times ranging from 5--30~s. We removed saturated data points using a square aperture centered around each target, check, and comparison star to ensure that no pixels had met or exceeded the expected saturation level.  If a pixel in the aperture met or exceeded this level, the data point corresponding to the image containing the object was removed from analysis.  Subsequent observations had shortened exposure times (15~s and 5~s) to avoid saturation.  All data was recorded in the FITS file format \citep{1981A&AS...44..363W} and processed by SIA. These results can be seen in Table~\ref{tab:siatable}. In addition to our own photometric measurements, we included observations from the AAVSO International Database (\href{https://www.aavso.org/aavso-international-database}{AID}; \citet{kafka_2020}) in our analysis.

%%%%%%%%%%%%%%%%%%%%%%%%%%%%%%%%%%%%%%%%%%%%%%%%%%%%%%%%%%%%

\section{Image Processing} \label{sec:improc}

We used our STEPUP Image Analysis code to process the photometric data taken by STEPUP of the 2018 outburst of AG~Dra. SIA works in three main steps: (1) instrument signature removal (ISR); (2) astrometric calibration (ASTROM); and (3) differential photometry (PHOT). As input, SIA takes raw science images in the FITS file format, three types of calibration images, a plate-solved science image \citep[generated by Astrometry.net][]{2010AJ....139.1782L}, and an input file that includes coordinates of the target, check, and comparison stars and the magnitudes of the comparison stars.  We list the comparison stars used for AG~Dra's analysis in Table~\ref{tab:compstars} given by the \emph{AAVSO Variable Star Plotter} tool. All three steps of SIA were performed to process the AG~Dra data and are summarized as follows. 
\par
For ISR, SIA writes a data set of files that have been corrected for dead pixel columns, uneven CCD illumination, and thermal noise using flat, bias, and dark calibration images. To generate master calibration files, SIA takes the median across the image set of each pixel for the dark and bias. For the master flat, the array is normalized with respect to the center region of the image that is evenly illuminated. The raw science images have the master bias and dark subtracted from them and are divided by the flat. The result is an instrument-signature-removed data set.
\par
\begin{deluxetable}{lrrrr}
\tablecaption{Comparison stars used to process AG~Dra data. \label{tab:compstars}}
\tablehead{
\colhead{AUID} & \colhead{Label} & \colhead{R.A.} & \colhead{Dec.} & \colhead{R} \\
 & & \colhead{[HH:MM:SS]} & \colhead{[DD:MM:SS]} & \colhead{[mag]}
}
\startdata
000-BCY-347 & 129 & 16:00:08.77 & 66:49:20.0 & 12.555 \\
000-BCY-346 & 123 & 16:00:24.08 & 66:49:29.6 & 11.980 \\
000-BJS-730 & 111 & 16:02:54.40 & 66:41:33.9 & 10.708 \\
000-BCY-344 & 119 & 16:00:11.22 & 66:39:14.2 & 11.575 \\
000-BCY-348 & 132 & 16:01:08.41 & 66:55:21.4 & 12.900 \\
\enddata
\tablecomments{Comparison stars used for photometric analysis of AG~Dra data. These stars were given by the \emph{AAVSO Variable Star Plotter} Photometry Table with VSP code {\bf X24880AIL}.}
\end{deluxetable}
In the next step, ASTROM, SIA takes the instrument signature removed files generated by ISR and a plate-solved image generated by Astrometry.net to write a set of files with the WCS FITS header keywords of the plate-solved image to the headers of the rest of the dataset. Then, SIA uses the WCSTools software package \citep{1997ASPC..125..249M} to adjust this information to accurately represent the coordinates of each pixel in each individual file. The result is a dataset with instrument signature-removed, plate-solved images.
\par
The final step of SIA is to perform differential aperture photometry. This places apertures at the positions of the target, check, and several comparison stars to get the sum of counts in the aperture for each object in every image of the dataset. A background rate per square pixel ($s_\textrm{bkgd}$) for the region of the image is determined by placing an annulus around the aperture and dividing its count sum by its area. The aperture and annulus sizes are as follows: $r_\textrm{aper}=4$\arcsec, $r_\textrm{in}=25$\arcsec, and $r_\textrm{out}=27$\arcsec. Subtracting the product of $s_\textrm{bkgd}$ and the area of the aperture ($A_\textrm{aper}$) from the aperture sum gives the net counts of the object. A 2D-Gaussian fit is applied for aperture centroiding. This process is used to get the net counts for all objects of interest in each image. The uncertainty in net counts for an object is given by
\begin{equation}
    N=\sqrt{S_*+s_\textrm{bkgd}A_\textrm{aper}}\label{eq:err}
\end{equation}
where $S_*$ is the net count value in the aperture around the object.
\par
The net count values are then calibrated to magnitudes using the relation,
\begin{equation}
m_*=m_c-2.5\log_{10}\left(\frac{S_*}{S_c}\right), \label{eq:mag}
\end{equation}
where $m_*$ and $S_*$ are the magnitude and counts of the target star, respectively and for the comparison star the same values given by $m_c$ and $S_c$. SIA outputs a light curve of the target and check star as well as output files giving magnitude values and net count values for both objects as well as unscaled light curves of comparison stars and a summary of aperture position corrections.
\par
%%%%%%%%%%%%%%%%%%%%%%%%%%%%%%%%%%%%%%%%%%%%%%%%%%%%%%%%%%%%
\begin{figure*}
\centering
% \textbf{2018 Outburst of AG~Draconis}
\plotone{{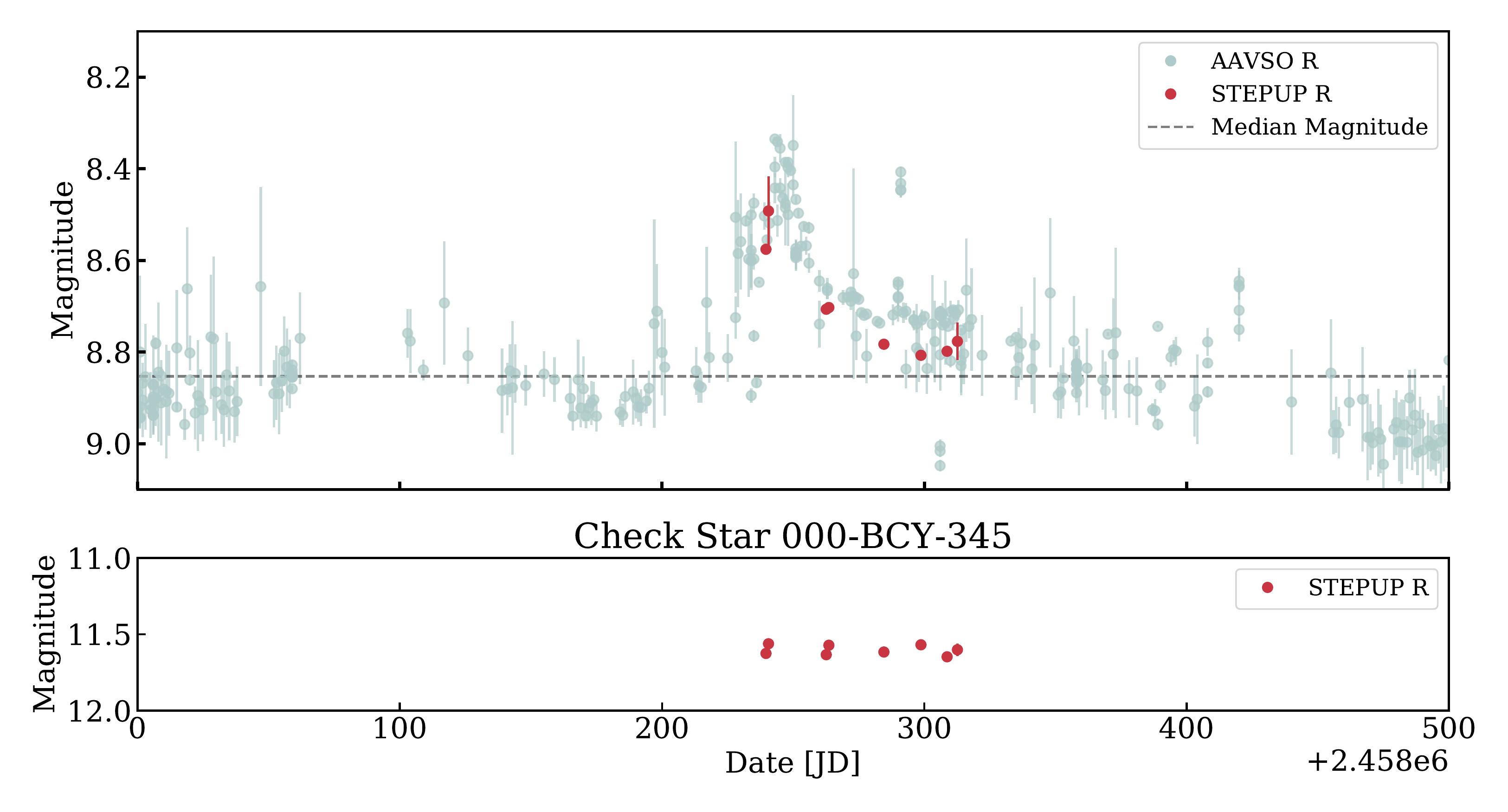}}
\caption{(Top) 2018 outburst of AG~Draconis shown by AAVSO and STEPUP R-band measurements. The system has an orbital period of $T_\textrm{orbit}=549.73~d$ and the cool component has a pulsation period of $T_\textrm{pulse}=355.27~\textrm{d}$ \citep{Galis1999}, which are visible during quiescence. (Bottom)  Light curve of check star 000-BCY-345. \label{fig:lceph}}
\end{figure*}

\begin{figure*}
\centering
% \textbf{Color Light Curve of AG~Dra 2018 Outburst}
\plotone{{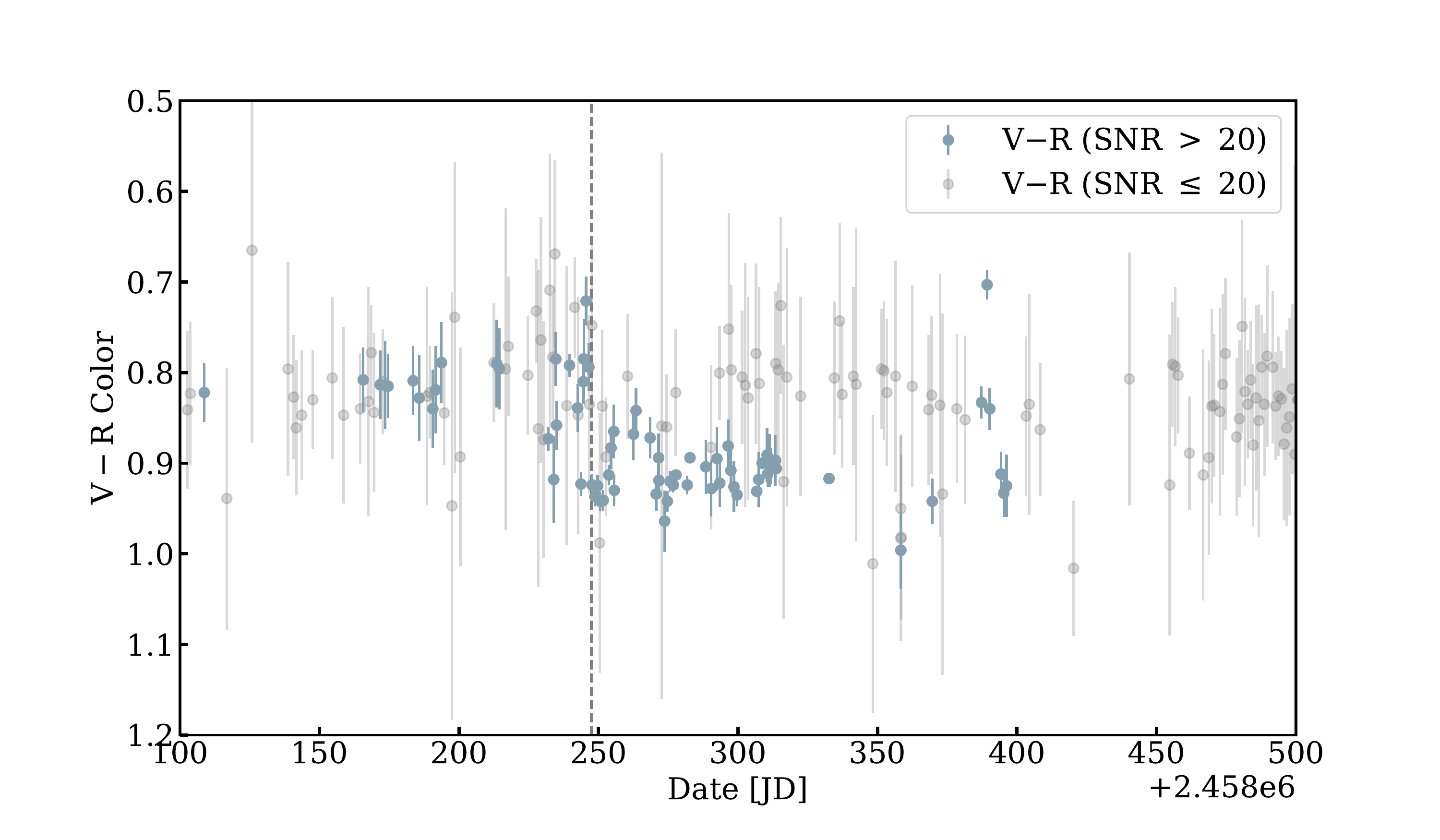}}
\caption{Color evolution of AG~Dra's 2018 outburst. The system's behavior exhibits chromaticity of amplitude $\textrm{V}-\textrm{R}=0.112 \pm 0.015~\textrm{mag}$. A slight increase in temperature can be seen leading up to the outburst's peak followed by a larger reddening as the system returns to quiescence. \label{fig:colorlc}}
\end{figure*}

\begin{figure*}
\centering
% \textbf{2018 Outburst of AG~Draconis}
\plotone{{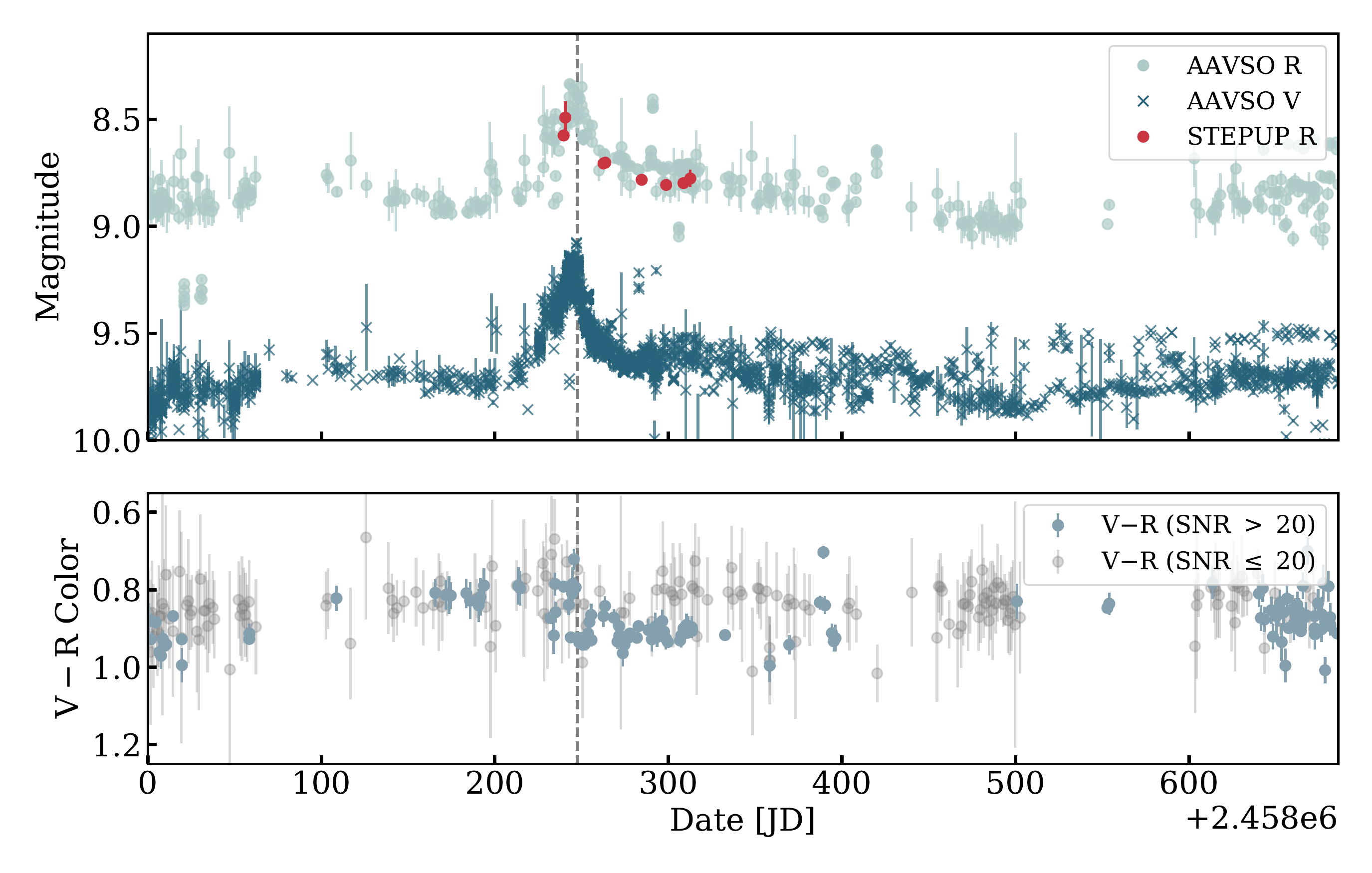}}
\caption{2018 outburst of AG~Draconis shown by STEPUP R-band and AAVSO R and V-band measurements. This shows all measurements available in the AID, instead of the median 1-d binned data as used in the analysis described in Section \ref{sec:analy}.  The R-band amplitude of the outburst was observed to be $\Delta\textrm{R}=0.518\pm0.011~\textrm{mag}$ and the V-band amplitude was observed to be $\Delta\textrm{V}=0.781\pm0.003~\textrm{mag}$. Below this is the $\textrm{V}-\textrm{R}$ light curve over the course of the outburst, showing an amplitude of 0.100~mag as the system returns to quiescence. The vertical dashed line marks the peak of the outburst, which occurred on JD~2458247.448. \label{fig:lc}}
\end{figure*}

\section{Analysis} \label{sec:analy}
We used data collected by STEPUP as well as R-band and V-band observations available from the AID to analyze AG~Draconis's 2018 outburst behavior. SIA was used to analyze each night of STEPUP data on the outburst and, assuming a Gaussian distribution, the median of observed magnitudes was taken to be the system's magnitude for a given night. The measurements for all eight nights of observation can be seen in Table~\ref{tab:obsresults} and a plot of STEPUP's measurements in Figure~\ref{fig:brightening}. We took the uncertainty in each night's magnitude, $\sigma_i$ to be the standard error of the data set,
\begin{equation}
    \sigma_i = \frac{\sigma}{\sqrt{n}},
\end{equation}
where $\sigma$ is the sample standard deviation from the median, and $n$ is the total number of data points. The light curve of the outburst including AAVSO and STEPUP data can be seen in Figure~\ref{fig:lceph}.
\par
We included AAVSO data to analyze the outburst depth and start and end date. The AID points used in our analysis are the median of all R-band observations in bins of 1~d. We took the quiescence value of AG~Draconis to be the median of these resulting magnitudes, giving a value of $\textrm{R}=8.853$~mag. Using this value as a threshold to distinguish between in-outburst and out-of-outburst data points, the outburst start and end dates are JD~2458190 and JD~2458351, respectively, giving an outburst duration of 161~d. Taking the difference of the minimum and maximum magnitude values during this period, we found an outburst depth of $\Delta\textrm{R}=0.518 \pm 0.011~\textrm{mag}$. Figure~\ref{fig:lc} presents the full outburst with AAVSO V and R observations as well as STEPUP R observations.
\par

\begin{deluxetable}{crrr}
\tablecaption{Observations of AG~Dra by STEPUP \label{tab:obsresults}}
\tablehead{
\colhead{Date} & \colhead{ExpTime} & \colhead{R} & \colhead{Uncertainty} \\
               & \colhead{[s]} & \colhead{[mag]} & \colhead{[mag]}
}
\centering
\startdata
2018-04-30 & 30 & 8.5762 & 0.0008 \\
2018-05-01 & 30 & 8.4682 & 0.0008 \\
2018-05-23 & 30 & 8.7039 & 0.0009 \\
2018-05-24 & 15 & 8.7393 & 0.0018 \\
2018-06-14 & 15 & 8.7904 & 0.0013 \\
2018-06-28 &  5 & 8.8538 & 0.0023 \\
2018-07-08 & 10 & 8.7884 & 0.0015 \\
2018-07-12 & 10 & 8.8002 & 0.0016 \\
\enddata
\tablecomments{Median magnitudes of AG~Dra for each night of observation by STEPUP. An outburst depth of $\Delta\textrm{R}=0.518 \pm 0.011~\textrm{mag}$ was observed using STEPUP and AAVSO measurements over a period of 161~d.}
\end{deluxetable}

To determine the nature of the outburst, we next performed a color analysis of AG~Dra's 2018 outburst. All STEPUP R and AAVSO V and R observations were used to give $\textrm{V}-\textrm{R}$ color during the outburst period. To get a higher-resolution light curve, instead of using 1-d bins as was used in the R-band analysis, we divided the light curves into intervals of 0.3~d ($\sim7.2~\textrm{h}$) where the median of all points in each interval was taken to be the value of that interval's magnitude. Each value's associated uncertainty was propagated to give the uncertainty in each interval's magnitude. Then, the color light curve and its uncertainty values were determined by subtracting the values in each band for each interval and propagating their uncertainties. To determine the amplitude of the color light curve all points in the outburst interval with $\textrm{SNR}>20$ were considered. By taking the difference in the median of the pre-outburst values and the post-outburst values we found a color of $\textrm{V}-\textrm{R}=0.112 \pm 0.015~\textrm{mag}$. This result can be seen in Figure~\ref{fig:colorlc}.

%%%%%%%%%%%%%%%%%%%%%%%%%%%%%%%%%%%%%%%%%%%%%%%%%%%%%%%%%%%%
\section{Results} \label{sec:res}
The AG~Draconis system was observed by STEPUP and AAVSO observers to outburst by $\Delta\textrm{R}=0.518$~mag over the course of 161~d, lasting from JD~2458190 until JD~2458351. The outburst peaked in the R-band on JD~2458242.749 and in the V-band on JD~2458247.448. This outburst exhibited color change of $\textrm{V}-\textrm{R}=0.112$~mag. This color change coincided with the V-band outburst's peak, so we will take JD~2458247.448 to be the date of the outburst's peak. The V-band depth of this outburst is similar to that of previous minor outbursts of AG~Dra, such as the system's 2016 outburst that peaked at around $\textrm{V}=9.1$~mag. \citet{Galis2017} studied this outburst by examining the system's equivalent widths of certain emission lines and the disappearance of the Raman scattered O VI lines. This study shows evidence of the outburst being of both hot and cool type. Our analysis of AG~Dra's 2018 outburst may suggest a similar temperature evolution, with the primary feature being a large reddening (and potentially a drop in temperature) following the outburst's peak, as shown in Figure~\ref{fig:lc}. Additionally, before AG~Dra began descending back to quiescence, a slightly bluer $\textrm{V}-\textrm{R}$ color can be seen as the system approaches its peak outburst value.
\par
There are three main pieces of evidence that suggest that this outburst was a disk instability: (1) the sharp increase in brightness followed by a longer descent to quiescence, (2) the scale of the outburst, and (3) the color evolution of the event. The system rose to outburst in 55.1~d. After a small amount of brightening for the first $\sim20$ days of the outburst, the system began to rapidly brighten, with its magnitude increasing linearly at a rate of $\textrm{V}=-0.018$~mag per day. A constraint on the timescale of this brightening can be seen in Figure~\ref{fig:brightening}, which shows no change in magnitude on the order of $\sim1.5$~hr. Following the outburst's peak, the system's brightness dropped off rapidly at first, declining at a rate of $\textrm{V}=0.033$~mag per day for the first $\sim15$~d, followed by a slower rate of decline for the duration of the system's return to quiescence. This exponential fall in brightness provides evidence against outbursts that typically have linear rates of decline, such as classical novae \citep{Hachisu_2015}. The system took $\sim105.9$~d to completely return to quiescence. It is quite typical of disk instability-driven dwarf novae to have brightening times that are shorter than the the timescale of their decline, as is seen in this outburst. This model would suggest that the system's brightness declines due to the propagation of a cooling wave inward through the disk at the local sound speed. \citet{2019MNRAS.487.2166L} provides an upper limit on the size of the WD's accretion disk of 65~$R_\odot$. For a rate of propagation of 0.7~$R_\odot$ per 7--10~d \citep{Sokoloski_2006}, this is reasonable, though it would suggest a much smaller disk size than the provided upper limit. 
\par
Furthermore, the amplitudes of $\Delta\textrm{R}=0.518~\textrm{mag}$ and $\Delta \textrm{V}=0.781~\textrm{mag}$ are too small to be caused by the thermonuclear runaways that drive classical symbiotic outbursts. Viewing the color evolution of the system, we see that the system became slightly bluer as the peak of the outburst occurred, followed by sizable reddening corresponding to the $\textrm{V}-\textrm{R}=0.112$~mag amplitude of the light curve after the outburst's peak. This provides evidence against classical novae since these types of outbursts usually show a negative color (i.e. $\textrm{B}-\textrm{V}$ color $< 0~\textrm{mag}$ and $\textrm{U}-\textrm{B}$ color $< 0~\textrm{mag}$, according to \citet{Hachisu_2015}) following the peak of the event. Given our unresolved photometric data of the entire system, it is impossible to know which component of AG~Dra was responsible for this increase in temperature. In the disk instability-type outburst, we see a rise in accretion disk temperature that triggers a change in the disk's viscosity as it reaches a critical temperature. This change in viscosity causes an increase in mass flow through the disk and subsequent heating and brightening, which could be responsible for the behavior of the color light curve as the outburst reaches its peak. Then, as the event ends, the system cools until its normal temperature is restored by the lower rate of mass flow supplied by the cool component's mass loss, which could in theory be responsible for the increase in $\textrm{V}-\textrm{R}$ color seen in Figure \ref{fig:colorlc}. If it were confirmed that this temperature change corresponds to a change in the disk temperature, then this would provide further evidence for the disk instability nature of this outburst.
\par
Another model that is less suited to describe this event is the combination nova outburst. The combination nova outburst is also triggered by disk instabilities, but is followed by a large decrease in temperature and increase in brightness as the white dwarf expels a surrounding shell of material after enhanced thermonuclear burning has commenced. This seems less likely to have caused this event, as the peak luminosity caused by enhanced shell burning would be much higher than that observed in AG~Dra's outburst. Also, a combination nova type outburst would most probably not have a linear rise to peak luminosity, as is seen in this event. While the available evidence favors the disk instability model, further data would be useful to distinguish between the temperatures of the disk, hot component, and cool component. In the case of a disk instability outburst, we would expect an increase in temperature and luminosity of the accretion disk, while the other components remain fixed in these parameters. While the shape and timescales of this 2018 event are generally consistent with those of typical dwarf novae, the expected linear decline corresponding to the propagation of a cooling wave through the disk is not visible. The system’s R-band and V-band brightness both fall off exponentially, indicating that there may be further activity involved in the system's cooling and decline in brightness. Additionally, if the size of the disk is as large as the upper limit provided by \citet{2019MNRAS.487.2166L}'s study, this cooling time would not be consistent with the cooling rate described by \citet{Sokoloski_2006}. Further data that resolves the activity of individual components of the system may be illuminating in consideration of this cooling mechanism.
%%%%%%%%%%%%%%%%%%%%%%%%%%%%%%%%%%%%%%%%%%%%%%%%%%%%%%%%%%%%

\section{Discussion} \label{sec:future}
Since the conclusion of this event, AG~Dra has not exhibited any further outbursts, with a notable lack of activity in 2019 May during the time when the next outburst of AG~Dra was expected. This suggests that the 2015-2018 active phase of the system has concluded. Though this active stage's outburst frequency has remained consistent with previous active stages, continuous UBVR photometric monitoring of the system is still necessary to determine if AG~Dra has truly returned to quiescence or if it will continue to exhibit abnormal outburst behavior. In particular, monitoring the temperature evolution of the hot component and accretion disk individually would be especially helpful in looking for signs of combination nova-type outbursts.
\par
 Though this event was probably triggered by disk instabilities, it remains unclear what caused the discrepancy between this outburst's exponential fall-off and the typical dwarf novae's linear decline. For the typical major outbursts exhibited by AG~Dra in its active phase the combination nova model shows strong potential of explaining the underlying mechanism for at least some of the outbursts, though it has not been confirmed as conclusively as in \citet{Sokoloski_2006}'s study of Z~And. What remains unclear about the system is the connection between the minor outbursts exhibited by AG~Dra in its 2015--2018 active phase and its typical behavior during major outbursts. Whether or not there is a connection between this activity and previous outbursts has yet to be determined. Knowing the temperature and individual luminosities of each component would clarify whether this is indicative of a different outburst mechanism (e.g., a combination nova-type outburst) or if this behavior is due to system properties of AG~Dra, such as having a small disk size or interference of thermal pulsations by the cool component, allowing us to connect this activity into the grand scheme of AG~Dra's outburst behavior.
%%%%%%%%%%%%%%%%%%%%%%%%%%%%%%%%%%%%%%%%%%%%%%%%%%%%%%%%%%%%
\acknowledgments
We acknowledge with thanks the variable star observations from the AAVSO International Database contributed by observers worldwide and used in this research. We also would like to thank both Scott Kenyon and the anonymous referee for their helpful comments that improved this paper.

This research was funded by the NASA Pennsylvania Space Grant Consortium Research Scholarship Award and the University of Pittsburgh Department of Physics \& Astronomy.
%%%%%%%%%%%%%%%%%%%%%%%%%%%%%%%%%%%%%%%%%%%%%%%%%%%%%%%%%%%%
\citestyle{aasjournal}
\bibliography{references}
%%%%%%%%%%%%%%%%%%%%%%%%%%%%%%%%%%%%%%%%%%%%%%%%%%%%%%%%%%%%

\end{document}